\documentclass{article}

\usepackage{amsmath}
\usepackage{amsfonts}
\usepackage{amssymb}
\usepackage{color}  
\usepackage{amsthm}
\usepackage{graphicx}
\usepackage{caption}
\usepackage{comment}
\usepackage{subcaption}
\usepackage{algpseudocode}
\usepackage{algorithm}
\usepackage[utf8]{inputenc}
\usepackage{tabularx}
\usepackage{multirow}
\usepackage{hyperref}
\usepackage[english]{babel}
\usepackage{setspace}

\newtheorem{theorem}{Theorem}

\usepackage{natbib}
\bibpunct[, ]{(}{)}{,}{a}{}{,}%

\title{Family Column Generation: A Principled Stabilized Column Generation Approach}

\author{Naveed Haghani\textsuperscript{\rm 1}, Julian Yarkony\textsuperscript{\rm 2}, Amelia Regan\textsuperscript{\rm 3} \\[2ex] 

\textsuperscript{\rm 1}University of Maryland, College Park, MD\\

\textsuperscript{\rm 2}Laminaar Optimization Research Group, La Jolla CA\\ 

\textsuperscript{\rm 3}University of California, Irvine, CA\\
}\date{March 2021}

\begin{document}

\maketitle
\begin{abstract}
We tackle the problem of accelerating column generation (CG) approaches to set cover formulations in operations research. At each iteration of CG we generate a dual solution that approximately solves the LP over all columns consisting of a subset of columns in the nascent set. We refer to this linear program (LP) as the Family Restricted Master Problem (FRMP), which provides a tighter bound on the master problem at each iteration of CG, while preserving efficient inference. For example, in the single source capacitated facility location problem (SSCFLP) the family of a column $l$ associated with facility $f$ and customer set $N_l$ contains the set of columns associated with $f$ and the customer set that lies in the power set of $N_l$. 
The solution to FRMP optimization is attacked with a coordinate ascent method in the dual. The generation of direction of travel corresponds to solving the restricted master problem over columns corresponding to the reduced lowest cost column in each family given specific dual variables based on the incumbent dual, and is easily generated without resolving complex pricing problems. We apply our algorithm to the SSCFLP and demonstrate improved performance over two relevant baselines.
\end{abstract}
\newpage
\section{Introduction}
\doublespacing
Expanded linear programming (LP) relaxations provide much tighter bounds than compact relaxations in many problem domains in  operations research \citep{lubbecke2005selected,desrosiers2005primer}, and more recently in computer vision \cite{yarkony2020data, FlexDOIArticle}. Column generation \citep{barnprice}(CG) is used to solve these expanded formulations. CG proceeds by relaxing the binary valued constraint in the  expanded integer linear programming formulation then constructing a sufficient set of the columns to exactly solve the LP relaxation.  CG achieves this by iterating between \textbf{(1)} solving the LP over a limited subset of the variables (called the nascent set) and \textbf{(2)} computing negative reduced cost columns. The LP is called the restricted master problem (RMP) while the mechanism to generate negative reduced cost columns is called pricing. The problem of generating these reduced cost columns is problem domain specific; but typically involves solving a small scale integer program, often via a dynamic program \citep{lubbecke2005selected}. CG terminates when no negative reduced cost columns exist, at which point the solution is provably optimal for the expanded LP relaxation.  

CG can be accelerated by dual stabilization methods \citep{ben2006dual,haghani2020integer,du1999stabilized,FlexDOIArticle,Pessoa2018Automation}. Stabilization methods select the dual solution to do pricing on in such a manner as to decrease the number of iterations of CG needed to solve the problem. 

In this paper we introduce a novel dual stabilization approach which we refer to as Family Column Generation (FCG). At each iteration of FCG we generate a dual solution, which solves the RMP over all columns consisting of a subset of columns in the nascent set. However this does not require additional expensive calls to pricing. A special LP solver is developed to solve this LP exploiting its structural properties. To use this LP solver we require an oracle that can generate the lowest reduced cost column over a subset of the elements in an existing column and obeying the structural properties or that existing column. Such optimization is far easier than standard pricing.  

We organize this document as follows. In Section \ref{CG_review} we review the classic column generation formulation of expanded set cover problems. In Section \ref{litRev} we review the existing literature in dual stabilization. In section \ref{Implementation} we discuss implementation details that are necessary for our exposition of the Family CG approach in section \ref{sec_family_RMP}. 
In Section \ref{exper} we provide experimental validation of our approach on  single source capacitated facility location (SSCFLP) \citep{diaz2002branch}. In Section \ref{FutureWork} we discuss an alternative construction of families of a column as well as our plans to apply this technique to other combinatorial optimization problems which are of significant current interest. Section \ref{conc} concludes this paper.  
%

\section{Column Generation Review}
\label{CG_review}

In this section we review the core concepts in Column Generation (CG) required to understand our Family Column Generation (FCG) algorithm. 
 
\subsection{Basic Column Generation}
\label{rev_CG}
We now consider the basic CG algorithm in the context of the following broad class of set cover problems.
We use $N$ to denote the set of items to be covered which we index by $u$.  We use $F$ to denote the set of facilities which we index by $f$.  We use $\Omega$ which we index by $l$ to denote the set of feasible pairs of subset of $N$, and a single member of $f$.  We use $a_{ul}\in \{0,1\},a_{fl} \in \{0,1\}$ to describe $\Omega$.  Here $a_{ul}=1$ IFF $l$ includes item $u$ and otherwise $a_{ul}=0$.   Similarly $a_{fl}=1$ IFF $l$ includes uses facility $f$ and otherwise $a_{fl}=0$.  We associate each $l\in \Omega$ with a cost $c_l$.  We frame set cover as the following integer linear program which enforces that each facility is used no more than once.  Here $\theta_l \in \{0,1\}$ is a binary variable used to describe the solution to set cover.  Here $\theta_l=1$ IFF $l$ is included in the solution and otherwise is zero.    
\begin{subequations}
\label{basicILP}
\begin{align}
\label{eq_basic_obj}
    \min_{\theta_l \in \{0,1\} }\sum_{l \in \Omega}c_l \theta_l\\
    \label{eq_basic_cover}
    \sum_{l \in \Omega}a_{ul}\theta_l \geq 1 \quad \forall u \in N\\
    \label{eq_basic_pack}
    \sum_{l \in \Omega}a_{fl}\theta_l \leq 1 \quad \forall f \in F
\end{align}
\end{subequations}
We now describe the components of  \eqref{basicILP}.  In \eqref{eq_basic_obj} we seek to minimize the total cost of elements in $\Omega$ selected.  In \eqref{eq_basic_cover} we enforce that each item is included in at least one column.  In most applications \eqref{eq_basic_cover} is tight for all $u \in N$ in any optimal solution \cite{barnprice}.  In \eqref{eq_basic_pack} we enforce that each facility is used no more than once.   
Often we may enforce that the total number of columns selected does not exceed a particular value.  So as to not over-complicate the write up we do not include this but not that $c_l$ can be offset by a constant for all $l$ to encourage/discourage the use of more columns.  

Problems of the form \eqref{basicILP} are common in the operations research literature and include the single source capacitated facility location problem (SSCFLP) \citep{diaz2002branch} and the  vehicle routing problem (VRP) \citep{costa2019,Desrochers1992}.  

Solving \eqref{basicILP} is NP-hard however efficient exact or approximate algorithms can be built using the column generation (CG) framework \citep{barnprice}.  CG solves the LP relaxation of \eqref{basicILP}, which in many applications well approximates \eqref{basicILP}.

CG operates as follows.  CG generates a sufficient subset of $\Omega$ denoted $\Omega_R$ to provably solve optimization over all $\Omega$.  This is done by iterativley solving the LP over $\Omega_R$ and adding columns with negative reduced cost to $\Omega_R$.  CG terminates when no column in $\Omega_R$ has negative reduced cost. We write the primal and dual LP relaxations below over $\Omega_R$ with dual variables denoted in [\ref{basicLDual}] by their associated  constraints. 
\begin{subequations}
\label{basicLPrimal}
\begin{align}
    \mbox{Primal Optimization:  }\min_{\theta \geq 0}\sum_{l \in \Omega_R}c_l \theta_l\\
    \sum_{l \in \Omega_R}a_{ul}\theta_l \geq 1 \quad \forall u \in N \quad [\pi_u]\\
    \sum_{l \in \Omega_R}a_{fl}\theta_l \leq 1 \quad \forall f \in F \quad [\pi_f]
\end{align}
\end{subequations}
\begin{subequations}
\label{basicLDual}
\begin{align}
    \mbox{Dual Optimization:  }\max_{\pi \geq 0}\sum_{u \in N}\pi_u-\sum_{f \in F}\pi_f\\
    c_l+\sum_{f \in F}a_{fl}\pi_f-\sum_{l \in \Omega}a_{ul}\pi_u \geq 0 \forall l \in \Omega_R, [\theta_l]
\end{align}
\end{subequations}
CG optimization is initialized with columns $\Omega_R$ describing a feasible solution to  \eqref{basicLPrimal}.  Next CG iterates between the following two steps until no column has negative reduced cost.  \textbf{(1)}  Solve \eqref{basicLPrimal} providing a primal and a dual solution. \textbf{(2)} Find the lowest reduced cost column associated with each $f \in F$, and add the negative reduced cost columns identified to $\Omega_R$.  This step is referred to as pricing.  CG terminates when pricing  fails to identify a negative reduced cost column.  The corresponding primal solution $\theta$ is provably optimal for optimization over $\Omega$ at termination of CG.  

Pricing is written below given fixed $f$ using $\Omega_f$ to denote the subset of columns associated with $f$.
\begin{subequations}
\label{pricer}
\begin{align}
    \min_{l \in \Omega_f}\bar{c}_l\\
    \bar{c}_l=c_l+\pi_f-\sum_{u \in N} a_{ul}\pi_u
\end{align}
\end{subequations}
Solving pricing is typically a combinatorial optimization problem that is problem domain specific.   Most commonly it is solved as a dynamic program or resource constrained shortest path problem \citep{lubbecke2005selected} though can be a small scale ILP \citep{FlexDOIArticle,zhang2017efficient}.   

 The solution at termination of optimization may still be fractional. Multiple mechanisms can be used to tackle this issue while achieving exact optimization.  The set cover LP relaxation can be further tightened in a cutting plane manner by valid inequalities such as subset-row inequalities \citep{jepsen2008subset,wang2017tracking}.  The use of such valid inequalities permits optimization with CG.  CG can be built into an exact branch-bound search procedure using branch-price \citep{barnprice}.

\section{Related Literature}
\label{litRev}

The literature on column generation techniques is vast. Here we review the most relevant material, namely research on trust region based methods and dual optimal inequalities used to accelerate CG. CG suffers from slow convergence when the number of items in a column becomes large. This tends to produce intermediate dual solutions which are sparse and do not share properties with those of known dual optimal solutions.  We now consider some methods designed to circumvent this difficulty.   

%
%
\subsection{Trust Region Based Methods}
Trust regions based methods discourage \citep{du1999stabilized} or prevent \citep{marsten1975boxstep} the next dual solution from leaving the area around the best dual solution generated thus far (in terms of Lagrangian relaxation). This is done since the current set of columns provides little information regarding the Lagrangian relaxation of dual solutions dis-similar to a previously generated dual solution. 

Smoothing based approaches \citep{Pessoa2018Automation} are a simple class of trust region approaches that achieve excellent results in practice. Smoothing based approaches only differ from standard CG in the selection of the dual variables terms which pricing is done on. Specifically they use a convex combination of:
\\
1) The dual solution of the current restricted master problem and \\
2) The dual solution generated thus far with greatest Lagrangian relaxation. 


\subsection{Dual Optimal Inequalities}
Dual Optimal Inequalities (DOI)\citep{ben2006dual} provide provable bounds on the optimal dual solution and ensure that the each step dual solution lies in the corresponding space. The primal DOI correspond to slack variables in the primal such as providing rewards for over-covering items or costs for swapping one item for another \citep{Gschwind2016Dual}. DOI are provably inactive at termination of CG but may or may not be active in intermediate steps.

\subsection{Flexible Dual Optimal Inequalities}
We now consider the Flexible Dual Optimal Inequalities (F-DOI) \citep{FlexDOIArticle,haghani2020relaxed,haghani2020smooth}. F-DOI exploit the following observation:  the change in the cost of a column induced by removing a small number of items is often small and can be easily bounded. Such bounds are column specific and exploit properties of the problem domain. In the primal form additional variables which provide reward for over-covering items are included. These rewards are set such that at optimality they are not used, but prior to termination of CG they have an important role. This can be understood as adding to the RMP all columns consisting of subsets of columns in the RMP. The costs of these columns provide upper bounds on the true cost of the column. F-DOI and their predecessors, varying and invariant DOI \citep{yarkony2020data}, provide considerable speed-ups and also make CG more robust to the specific selection of optimization parameters.   

\section{Implementation Details}
\label{Implementation}
Here we review Lagrangian Relaxation and our test problem, the single source capacitated facility location problem. Further, we review the box-step method because we will need it in our FCG approach.

\subsection{Lagrangian Relaxation}
\label{rev_Lag}
 The Lagrangian relaxation \citep{desrosiers2005primer} is smooth, convex and has identical value to the MP at the optimizing $\pi$ for the MP. 
At any given point in CG optimization a lower bound on the optimal solution can be generated using the Lagrangian relaxation. The Lagrangian relaxation can be used to provide as stopping criteria for CG. Specifically when the difference between the objective of the RMP solution and the Lagrangian relaxation is sufficiently small (according to a user defined criteria) we terminate CG optimization. We write the Lagrangian relaxation at a given dual solution $\pi$ as $\ell_{\pi}$, which we define below.  
\begin{subequations}
\label{ellDef}
\begin{align}
\label{ellDefBasic}
    \mbox{MP}\geq \ell_{\pi}=\sum_{u \in N} \pi_u -\sum_{f \in F}\pi_f+\sum_{f \in F}  \min(0,\min_{l \in \Omega_{f}}\bar{c}_l)
\end{align}
\end{subequations}
Observe that any non-negative solution $\pi$ can be projected to one that is a feasible solution to the dual master problem  (\eqref{basicLDual} with $\Omega_R \leftarrow \Omega$) with objective identical to the Lagrangian relaxation in \eqref{ellDefBasic} by setting $\pi_f\leftarrow \pi_f+ \min(0,\min_{l \in \Omega_{f}}\bar{c}_l)$.
%
  %
\subsection{Application: Single Source Capacitated Facility Location}
\label{rev_sscflp}
To provide a meaningful example we consider the classical single source capacitated facility location problem (SSCFLP) which is often used to explore column generation techniques \citep{diaz2002branch,haghani2020smooth}.

In the SSCFLP, we are given a set of customers $N$ and a set of potential facilities $F$, such that $F\cap N=\emptyset$. Each facility $f\in F$ is associated a fixed opening cost $c_f$ and a capacity $K_f$. Each customer $u\in N$ is associated a demand $d_u\geq 0$. Each pair $(f, u)\in F\times N$ is associated an assignment cost $c_{fu}\geq 0$.  Without loss of generality, we may assume that all parameters are integer-valued. The SSCFLP consists in selecting a subset of facilities to open, and to assign every customer to exactly one open facility in such a way that the capacities of the selected facilities are respected, at minimum total cost (fixed costs + assignment costs). 

The pricing subproblem is a 0-1 knapsack problem which is easy to solve in practice\citep{cuttingstock,gilmore1965multistage}. Specifically we condition on facility $f$ and solve the knapsack problem below. We use $x_{u} \in \{0,1\}$ to denote the decision variable for customer $u$; where $x_u=1$ IFF customer $u$ is included.  
\begin{subequations}
\label{pricingKnap}
\begin{align}
\min_{x_u \in \{0,1\} \quad \forall u \in N}c_f+\pi_f+\sum_{u \in N}x_u(c_{fu}-\pi_u)\\
\sum_{u \in N}x_u d_u \leq K_f
\end{align}
\end{subequations}
%
\subsection{Box-Step Method for CG Optimization}
\label{rev_box}
  The box-step method is a classic mechanism to accelerate CG \citep{marsten1975boxstep} which lies in the family of trust region methods \citep{du1999stabilized}. Trust region methods broadly can be understood from the dual perspective as follows. The current set of columns $\Omega_R$ provide a good estimate of the Lagrangian relaxation for dual solutions $\pi$ that are very similar to previously generated dual solutions over the course of CG optimization. However $\Omega_R$ does not provide a good estimate of the Lagrangian relaxation for highly distinct points, thus grossly overestimating the dual objective for solutions which are dissimilar to previous solutions. Trust region based methods ensure or encourage the new solution to lie near the previously generated optimal dual solution. Trust region methods thus make CG approaches operate more like gradient ascent methods. By this we mean that that $\Omega_R$ and the dual solution change gradually based on each-other. We now describe a trivial variant of the box-step.  
%
  %
   For efficiency of notation we use $\Psi_{\pi^+,\pi^-,\hat{\Omega}}$ to be the value of the LP over the box defined by  $\pi^+$ and $\pi^-$ over a set $\hat{\Omega}\subseteq \Omega$.%
\begin{subequations}
\label{boundRMP}
\begin{align}
    \Psi_{\pi^+,\pi^-,\hat{\Omega}}=\max_{\substack{\pi \geq 0 }} \sum_{u \in N} \pi_u-\sum_{f \in F}\pi_f
   \\
    c_l+\sum_{f \in F}a_{fl}\pi_f-\sum_{u \in N}a_{ul}\pi_u \geq 0 \ \quad \forall l \in \hat{\Omega}\\
    \label{boundPI}
    \pi^+_u\geq \pi_u \geq \pi^-_u \quad \forall u \in N
\end{align}
\end{subequations}
Note that \eqref{boundRMP} always has a dual feasible solution since $\pi_f$ is not restricted.  

The primal form of optimization in  \eqref{boundRMP} is written as follows using $\delta^+_u,\delta^-_u$ to denote the primal variables associated with \eqref{boundPI}.  So as to avoid confusion with the dual problem we use $\Delta^+_u\leftarrow \pi^+_u$ and $\Delta^-_u\leftarrow \max(0,\pi^-_u)$.
\begin{subequations}
\label{primal_box}
\begin{align}
\min_{\substack{\theta \geq 0\\ \delta \geq 0}}\sum_{l \in \hat{\Omega}} c_l \theta_l+\sum_{u \in N}\Delta^+_u\delta^+_u-\Delta^-_u\delta^-_u\\
\delta_u^+-\delta_{u}^- +\sum_{l \in \hat{\Omega}}a_{ul}\theta_l\geq 1 \quad \forall u \in N\\
\sum_{l \in \hat{\Omega}}a_{fl}\theta_l\leq 1 \quad \forall f \in F
\end{align}
\end{subequations}
The trivial box-step method proceeds iterates between the following two steps. \textbf{(1)}  We solve  \eqref{boundRMP} around the solution $\pi^*$ which is the dual solution with highest Lagrangian relaxation generated thus far. The box around $\pi^*$ is parameterized by a fixed constant $\nu \in \mathbb{R}_{+}$; where $\pi^+_u\leftarrow \pi^*_u +\nu$ and $\pi^-_u\leftarrow \max(0,\pi^*_u -\nu)$ for all $u \in N$.  (\textbf{2})is performed as in standard CG. We iterate between solving \eqref{primal_box}, and pricing until no column has negative reduced cost and the $\delta$ terms are inactive.  This certifies that an optimal solution has been produced to the master problem (MP).  The selection of $\nu$ can be understood in terms of the following trade off.  Smaller values of $\nu$ means each iteration of box-step makes less progress towards solving the MP.  However if $\nu$ is smaller then any given iteration is more inclined to improve the Lagrangian relaxation and hence update $\pi^*$. Sophisticated box-step methods can employ schedules and mechanisms for changing $\nu$ over the course of optimization.   
We write the box-step method formally in Alg \ref{basicBOx}, which we annotate below.
\begin{algorithm}[!b]
 \caption{Box-Step Method}
\begin{algorithmic}[1] 
\State $\Omega_R,\nu,\pi^* \leftarrow $ from user
\label{Zline_rec_input_start}
\Repeat
\label{Zline_outer_start}
%
 \State $\Delta^+_u, \leftarrow \pi^*_u+\nu$ for all $u \in N$
\label{Zset_delta_1}
\State $\Delta^-_u,\leftarrow \max(0,\pi^*_u -\nu)$ for all $u \in N$
\label{Zset_delta_2}
\State  Solve for $\theta,\bar{\pi},\delta$  using \eqref{primal_box} over $\Omega_R$
\label{Zline_solve_FRMP}
 \For{$f \in F$}
 \label{Zline_pricing_Start}
 \State $l^*_f\leftarrow \mbox{arg}\min_{l \in \Omega_f} c_l-\bar{\pi}_f-\sum_{u \in N}\bar{\pi}_u a_{ul}$
 \If {$0>c_{l^*_f}+\bar{\pi}_f-\sum_{u \in N}\bar{\pi}_u a_{ul^*_f}$}
 \State $\Omega_R \leftarrow \Omega_R \cup l^*_f$
 \EndIf
 \EndFor
  \label{Zline_pricing_End}
 \If{$\ell_{\pi^*}\leq \ell_{\bar{\pi}}$}
  \label{Zstore_best_start}
 \State $\pi^*\leftarrow \bar{\pi}$ 
 \EndIf
  \label{Zstore_best_end}
 \Until{$c_{l^*_f}+\bar{\pi}_f-\sum_{u \in N}\bar{\pi}_u a_{ul^*_f} \geq 0$ for all $f \in F$ and $\delta^+_u=\delta^-_u=0$ for all $u \in N$}
  \label{Zline_outer_end}
 \State Return last $\theta$  generated.  \label{ZreturnSol}
\end{algorithmic}
\label{basicBOx}
\end{algorithm}
%
%
\begin{itemize}
    \item Line \ref{Zline_rec_input_start}:  We receive the input $\Omega_R$ which provides for a feasible solution for the MP.  We also receive step size $\nu$ and initial $\pi^*$  which can be set trivially to the RMP solution over $\Omega_R$ or any other mechanism such as the zero vector.  
    \item Lines \ref{Zline_outer_start}-\ref{Zline_outer_end}:  Solve the MP over $\Omega$.  We certify that we have solved optimization by terminating when no column has negative reduced cost and $\delta$ is zero valued. We should note that if $\Delta^-_u=0$ then we do not include the primal variable  $\delta^-_u$ as the corresponding dual constraint is redundant.  We thus set $\delta^-_u=0$ by force when $\Delta^-_u=0$.   
    \begin{enumerate}
    \item Line \ref{Zline_solve_FRMP}:  Receive  solution to RMP over $\Omega_R$ in box around $\pi^*$. 
    \item  Lines \ref{Zline_pricing_Start}-\ref{Zline_pricing_End}:  Compute the lowest reduced cost column associated with each $f \in F$.  Then add any negative reduced cost columns computed to $\Omega_R$. 
    \item Lines \ref{Zstore_best_start}-\ref{Zstore_best_end}:  We store the best solution found thus far.  Here best means the one which maximizes the Lagrangian relaxation. 
    \end{enumerate}

    \item Line\ref{ZreturnSol}: Return the last solution generated $\theta$ which is provably optimal and feasible for the MP.  
\end{itemize}

\section{Family Column Generation}
\label{sec_family_RMP}
In this section we introduce our Family Column Generation (FCG) algorithm. FCG differs from standard CG only in the mechanism to generate dual solutions, upon which pricing is performed. FCG (approximately) solves the Family Restricted Master Problem (FRMP) at each at iteration of CG; where the FRMP better approximates the  MP at each iteration of CG than the standard RMP, while preserving efficient inference. The FRMP describes the LP over the set of all columns that lie in the family of any member of $l \in \Omega_R$, which we denote $\Omega^+_R$.  We define $\Omega_{\hat{l}}$ to be the family of columns corresponding to subsets of the items that compose $l$, denoted $N_{\hat{l}}$ ($N_{\hat{l}}=\{u \in N \mbox{, s.t. } a_{ul}=1\}$) and preserve structural properties of $l$ (such as $f_l$ and ordering of items if an ordering exists). Thus $\Omega_R^+=\cup_{\hat{l} \in \Omega_R}\cup_{l \in \Omega_{\hat{l}}}$.

We now describe $\Omega_{\hat{l}}$ for two applications. \\

Consider in SSCFLP a column $\hat{l}$ that is associated with facility $f$ and items $N_{\hat{l}}$. The set $\Omega_{\hat{l}}$ is the set of columns each associated with facility $f$ and a customer set that lies in the power set of $N_{\hat{l}}$. \\

In the case of capacitated vehicle routing problem (CVRP) 
$\Omega_{\hat{l}}$ contains the set of columns in which the order of customers visited is the same as $l$ but a subset (perhaps empty) of those customers are removed.  

The solution to FRMP optimization is attacked with a coordinate ascent method in the dual. In this method we generate a direction of travel then travel in that direction the optimal amount. We repeat this until we no longer improve the FRMP objective. FCG terminates only when FRMP is solved exactly and no column has negative reduced cost certifying that the solution is optimal for the MP.  

We organize this section as follows. In Section \ref{genDirec} we consider the generation of directions of travel. In Section \ref{optTrav} we consider the determination of the optimal travel distance. In Section \ref{sec_algForm} formalize FCG in algorithmic form and provide analysis. In Section \ref{convergAnal} we show that FCG optimally solves the MP. In Section \ref{speedInner} we accelerate FCG by permitting early termination of FRMP solution without compromising convergence of FCG to the optimal solution of the MP. In Section \ref{applicationSSCFLP} we discuss the use of FCG for SSCFLP.
\subsection{Generation of Direction of Travel}
\label{genDirec}
We now consider the generation of directions to improve an incumbent dual solution denoted $\pi^0$. This solution is based on creating an approximation to the dual FRMP around $\pi^0$ where $\nu$ describes the size of the window over which our approximation is constructed.  The value $\nu$ trades off the size of the space the approximation is done over and the quality of the approximation. Thus $\nu+\pi^0_u=\pi_u^+\geq \pi_u \geq \pi^-_u=\max(0,\pi^0_u-\nu)$ for all $u \in N$ and $\pi_f$ unrestricted. We then optimize an upper bound on the MP over this local approximation providing a solution $\bar{\pi}$. Then we travel from $\pi^0$ in the direction of $\vec{\pi}$ where $\vec{\pi}\leftarrow \bar{\pi}-\pi^0$. 

We seek to construct a set $\Omega_{R\pi^+}$ so as to provide a good approximation to $\Psi_{\pi^+,\pi^-,\Omega^+_R}$ that is no worse than that of $\Psi_{\pi^+,\pi^-,\Omega_R}$ at any point in the window of ($\pi^+,\pi^-$). Thus the following should be satisfied.   $\Psi_{\pi^+,\pi^-,\Omega_R}\geq \Psi_{\pi^+,\pi^-,\Omega_{R\pi^+}}\approx \Psi_{\pi^+,\pi^-,\Omega^+_R} $. 

Since we have to construct many approximations in our procedure to solve FRMP; the construction of $\Omega_{R\pi^+}$ must not use expensive operations such as calls to non-trivial pricing problems, or  calls to an LP solver, and must not result in an explosion in the number of columns. 
Our construction is motivated by the following equivalent way of writing optimization over $\Omega^+_R$ over the  window defined by $\pi^+,\pi^-$.

\begin{subequations}
\label{dual_FMP2}
\begin{align}
\max_{\substack{\pi \geq 0 }} \sum_{u \in N} \pi_u-\sum_{f \in F}\pi_f
   \\
   \label{minTerm}
    \min_{l \in \Omega_{\hat{l}}}c_l+\sum_{f \in F}a_{fl}\pi_f-\sum_{u \in N}a_{ul}\pi_u \geq 0 \quad \forall \hat{l} \in \Omega_R\\
    \pi^+_u \geq \pi_u \geq \pi^-_u \quad \forall u \in N
    \end{align}
\end{subequations} 

Optimization over \eqref{dual_FMP2} is intractable but can be efficiently approximated using the following upper bound. We replace the $\pi_u$ terms in \eqref{minTerm} with $\pi^+_u$. Thus there is only one constraint for each $l \in \Omega_R$. Furthermore in many applications solving the optimization in \eqref{minTerm} is as easy as shown in our application; No expensive calls to NP-hard pricing oracles are required. Using \eqref{minTerm} we define $\Omega_{R\pi^+}$ below.  

\begin{subequations}
\label{projSet}
\begin{align}
    \Omega_{R\pi^+}=\cup_{\hat{l} \in \Omega_R}\hat{l}_{\pi^+}\\
    \hat{l}_{\pi^+}= \mbox{arg}\min_{l \in \Omega_{\hat{l}}}c_{l}+\sum_{f \in F}a_{fl}\pi_f-\sum_{u \in N_{\hat{l}}}a_{ul}\pi^+_u
    \label{projection_eq}
\end{align}
\end{subequations}

We now show that $\Psi_{\pi^+,\pi^-,\Omega_R}\geq \Psi_{\pi^+,\pi^-,\Omega_{R\pi^+}}$

\begin{theorem}

Consider that the claim is false.  Thus there must exist a solution $\pi$ which is feasible for the optimization over  $\Psi_{\pi^+,\pi^-,\Omega_{R\pi^+}}$ but not for the optimization over  $\Psi_{\pi^+,\pi^-,\Omega_R}$ since the objective function of the two functions is identical.  There must exist a constraint $l\in \Omega_R$ for which the following holds.
%
\begin{align}
\label{contradictEQ}
-c_{\hat{l}}-\sum_{f \in F}a_{f\hat{l}}\pi_f+\sum_{u \in N} a_{u\hat{l}}\pi_u> -c_{\hat{l}_{\pi^+}}-\sum_{f \in F}a_{f\hat{l}_{\pi^+}}\pi_f+\sum_{u \in N} a_{u\hat{l}_{\pi^+}}\pi_u 
\end{align}
By construction in  \eqref{projection_eq} we know that for any $\hat{l} \in \Omega_R$ 
\begin{align}
\label{boundRQ}
    c_{\hat{l}}+\sum_{f \in F}a_{f\hat{l}}\pi^+_f-\sum_{u \in N_{\hat{l}}}a_{ul}\pi^+_u\geq c_{\hat{l}_{\pi^+}}+\sum_{f \in F}a_{f\hat{l}_{\pi^+}}\pi^+_f-\sum_{u \in N_{\hat{l}}}a_{u\hat{l}_{\pi^+}}\pi^+_u
\end{align}
Now we add \eqref{contradictEQ} and \eqref{boundRQ} creating the following inequality; which we then simplify on the succeeding line.
\begin{subequations}
\begin{align}
    \sum_{u \in N}a_{u\hat{l}}(\pi_u-\pi^+_u)> \sum_{u \in N}a_{u \hat{l}_{\pi^+}}(\pi_u-\pi^+_u)\\
    \label{mySumEQ}
    \sum_{u \in N}(a_{u\hat{l}}-a_{u \hat{l}_{\pi^+}})(\pi_u-\pi^+_u)> 0 
\end{align}
\end{subequations}
Observe that in \eqref{mySumEQ} that $(a_{u\hat{l}}-a_{u\hat{l}_{\pi^+}})$ is strictly non-negative since $ \hat{l}_{\pi^+}$ lies in $\Omega_{\hat{l}}$.  Observe also that  $(\pi_u-\pi^+_u)$ is strictly non-positive since $\pi^+_u\geq \pi_u$.  Thus $(a_{u\hat{l}}-a_{u\hat{l}_{\pi^+}})(\pi_u-\pi^+_u)$ is non-positive for every $u \in N$.  The sum of non-positive terms on the LHS of \eqref{mySumEQ} is thus non-positive.  Thus we have created a contradiction and hence proved the claim. \textbf{QED}.
\end{theorem}
Thus in this section we have provided a mechanism to produce directions of travel that locally approximate the FRMP over the area around an incumbent solution $\pi^0$.  %
%
%
%
%
%
%
\subsection{Determining the Optimal Travel Distance}
\label{optTrav}
In this section we determine the optimal distance to travel along the ray  starting at $\pi^0$ and traveling in direction $\vec{\pi}$ (where the ray is denoted $(\pi^0,\vec{\pi})$) so as to approximately maximize the Lagrangian relaxation (as written in \eqref{ellDefBasic}).  
%
%
%
%
Since evaluating the Lagrangian relaxation  requires a call to pricing, which may be expensive, we use a convex approximation to  \eqref{ellDefBasic} that like the Lagrangian relaxation is equal to the MP at an optimizing $\pi$ (for MP). This bound denoted $\ell^{\Omega^+_{R}}_{\pi}$ considers only columns in $\Omega^+_R$ and uses \eqref{projection_eq}, which is easy to compute. Below we define $\ell^{\Omega^+_R}_{\pi}$ as follows using helper term $\ell^{\hat{l}}_{\pi}$, which is the reduced cost of the lowest reduced cost column in the family of $\hat{l}$.   
%
\begin{subequations}
\label{ellDef2}
\begin{align}
    \label{ellDefADV}
    \ell_{\pi} \leq \ell^{\Omega^+_R}_{\pi}=\sum_{u \in N} \pi_u-\sum_{f \in F}\pi_f+\sum_{f \in F}\min_{\hat{l} \in \Omega_R \cap \Omega_f} \min(0,\ell^{\hat{l}}_{\pi})\\
    \ell^{\hat{l}}_{\pi}=\min_{l \in \Omega_{\hat{l}}}\bar{c}_l
\end{align}
\end{subequations}
%
Since $\pi$ is non-negative the maximum travel distance $m$ satisfies the following inequality.  
\begin{align}
\label{maxMval}
    m\leq \frac{-\vec{\pi}_z}{\pi^0_z} \quad \forall z \in N \cup F, \vec{\pi}_z <0
\end{align}
We set $m$ to the max possible value as described by \eqref{maxMval} unless it is unrestricted by \eqref{maxMval} and hence we set it to a large positive number ($m \leftarrow 100$ in experiments).
For any $\eta \in [0,1]$ we use $\phi_{\eta}$ to denote  $\ell^{\Omega^+_{R}}_{\pi+\eta m\vec{\pi}}$).  Since $\ell^{\hat{\Omega}}_{\pi}$ is a concave function of $\pi$ for any $\hat{\Omega} \subseteq \Omega$ we use the following binary search style procedure  to  maximize $\phi_{\eta}$ with respect to $\eta$.  At any given point in our procedure we preserve the following invariant $0\leq \eta^-\leq \eta^*\leq \eta^+\leq 1$ where $\eta^*=\mbox{arg}\max_{\eta \in [0,1]}\phi_{\eta}$.

We use $\zeta^-\leftarrow\frac{3\eta^-}{4}+\frac{\eta^+}{4},\zeta \leftarrow \frac{2\eta^+}{4}+\frac{2\eta^+}{4},\zeta^+\leftarrow\frac{1\eta^-}{4}+\frac{3\eta^+}{4}$ to be the one quarter,one  half and three quarter way points between $\eta^-$ and $\eta^+$.  At any given step of the iteration we evaluate $\phi_{\zeta^-},\phi_{\zeta},\phi_{\zeta^+}$.  We then update $\nu^+$,$\nu^-$ so as to preserve the invariant $ \eta^-\leq \eta^*\leq \eta^+$ while decreasing $\eta^+-\eta^-$. At any given point in optimization  there are three cases which describe the corresponding property of $\eta^*$.  
\begin{enumerate}
\item $\phi_{\zeta}=\max(\phi_{\zeta},\phi_{\zeta^+},\phi_{\zeta^-})$:  Then $\eta^*$  lies in $[\zeta^-,\zeta^+]$
    \item $\phi_{\zeta^-}=\max(\phi_{\zeta},\phi_{\zeta^+},\phi_{\zeta^-})$:  Then $\eta^*$  lies in $[\eta^-,\zeta]$
     \item $\phi_{\zeta^+}=\max(\phi_{\zeta},\phi_{\zeta^+},\phi_{\zeta^-})$:  Then $\eta^*$  lies in $[\zeta,\eta^+]$
\end{enumerate}
We formalize the search in Alg \ref{searchProc}.  
We terminate when we are within a user defined tolerance  $\epsilon$ ($\epsilon=10^{-5}$ in our experiments) then return the best point found thus far. 
\begin{algorithm}[!b]
 \caption{Optimization to determine $\eta^*$}
\begin{algorithmic}[1] 
\State $\eta^-\leftarrow 0$
\State $\eta^+ \leftarrow 1$
\While {$\eta^+-\eta^->\epsilon$}
\State $\zeta \leftarrow \frac{2\eta^-}{4}+\frac{2\eta^+}{4}$
\State $\zeta^+\leftarrow \frac{1\eta^-}{4}+\frac{3\eta^+}{4}$
\State $\zeta^-\leftarrow \frac{3\eta^-}{4}+\frac{\eta^+}{4}$
\If{$\phi_{\zeta}=\max(\phi_{\zeta},\phi_{\zeta^+},\phi_{\zeta^-})$}
\State $\eta^- \leftarrow \zeta^-$
\State $\eta^+ \leftarrow \zeta^+$
\State Continue
\EndIf
\If{$\phi_{\zeta^-}=\max(\phi_{\zeta},\phi_{\zeta^+},\phi_{\zeta^-})$}
\State $\eta^+\leftarrow \zeta $
\State Continue
\EndIf
\If{$\phi_{\zeta^+}=\max(\phi_{\zeta},\phi_{\zeta^+},\phi_{\zeta^-})$}
\State $\eta^-\leftarrow \zeta $
\State Continue
\EndIf
\EndWhile
\State Return $\mbox{arg}\max_{\eta \in \{ \eta^-,\eta^+\}}\phi_{\eta}$
\end{algorithmic}
\label{searchProc}
\end{algorithm}
%
\subsection{Algorithmic Formulation of Family-CG}
\label{sec_algForm}
We now consider an algorithmic formulation which we refer to as Family Column Generation (FCG). At each iteration of FCG we solve the Family RMP approximately then do pricing. The FCG is parameterized by one parameter only which is the step size $\nu$. The solution of the FRMP is found using the coordinate ascent approach of alternating between generating directions and going the optimal amount in that direction as measured by an approximation to the Lagrangian relaxation.  
 We terminate FRMP optimization when the current solution $\pi$ satisfies one of the following:
 \\ \textbf{(1)} $\pi$ describes the optimal solution to the RMP over $\Omega^+_R$ or
\\ \textbf{(2)} If the optimal distance to travel does not equal or exceed the minimum amount $\frac{1}{m}$. In Section \ref{convergAnal} we show that to ensure converge of FCG we must travel at least $\frac{1}{m}$ times the maximum possible distance.  

\begin{itemize}
    \item Line \ref{line_rec_input_start}:  We receive the input $\Omega_R$ which provides for a feasible though not optimal solution for the MP. We also receive step size $\nu$ and initial $\pi^*$  which can be set trivially to the RMP solution over $\Omega_R$ or any other mechanism such as the zero vector.  
    \item Line \ref{line_outer_start}-\ref{line_outer_end}:  Solve the MP over $\Omega$. We certify that we have solved optimization by terminating when no column has negative reduced cost.  Termination can only happen in this algorithm if FRMP is solved optimally.  
    \begin{enumerate}
    \item Line \ref{line_solve_FRMP}: Receive approximate solution to FRMP over $\Omega_R$.
    \item  Line \ref{line_pricing_Start}-\ref{line_pricing_End}:  Compute the lowest reduced cost column associated with each $f \in F$.  Then add any negative reduced cost columns computed to $\Omega_R$. 
    \item Line \ref{store_best_start}-\ref{store_best_end}:  We store the best solution found thus far.  Here best means the one which maximizes the Lagrangian relaxation. 
    \end{enumerate}

    \item Line \ref{returnSol}:Return the last solution generated $\theta$ which is provably optimal and feasible for the MP.  
\end{itemize}



\begin{algorithm}[!b]
 \caption{Family Column Generation}
\begin{algorithmic}[1] 
\State $\Omega_R,\nu,\pi^* \leftarrow $ from user
\label{line_rec_input_start}
\Repeat
\label{line_outer_start}

\State $\bar{\pi},\theta \leftarrow $ Solve FRMP approximately using Alg \ref{ezVer_master}  given $\Omega_R,\pi^*$ and $\nu$.
\label{line_solve_FRMP}
 \For{$f \in F$}
 \label{line_pricing_Start}
 \State $l^*_f\leftarrow \mbox{arg}\min_{l \in \Omega_f} c_l+\bar{\pi}_f-\sum_{u \in N}\bar{\pi}_u a_{ul}$
 \If {$0>c_{l^*_f}+\bar{\pi}_f-\sum_{u \in N}\bar{\pi}_u a_{ul^*_f}$}
 \State $\Omega_R \leftarrow \Omega_R \cup l^*_f$
 \EndIf
 \EndFor
  \label{line_pricing_End}

 \If{$\ell^{\Omega_R}_{\pi^*}\leq \ell^{\Omega_R}_{\bar{\pi}}$}
  \label{store_best_start}

 \State $\pi^*\leftarrow \bar{\pi}$ 
 \EndIf
  \label{store_best_end}
 \Until{$c_{l^*_f}+\bar{\pi}_f-\sum_{u \in N}\bar{\pi}_u a_{ul^*_f} \geq 0$ for all $f \in F$}
  \label{line_outer_end}
 \State Return $\theta$ for last $\theta$ generated.  \label{returnSol}
\end{algorithmic}
\label{ezVer}
\end{algorithm}

We now consider the solution to the FRMP problem.

\begin{itemize}
    \item Line \ref{line_rec_input_start2}: Receive input feasible solution, step size and initial solution $\pi^0$
    \item Line \ref{step_inner_start}-\ref{step_inner_end}: Solve the FRMP approximately using coordinate ascent. Terminate when either FRMP is optimally solved or there exists an $l \in \Omega^+_R-\Omega_R$ with negative reduced cost meaning that the updated step did not improve the Lagrangian relaxation over $\Omega^+_R$.   
    \begin{algorithm}[!b]
 \caption{Solve Family RMP}
\begin{algorithmic}[1] 
\State 
$\Omega_R,\nu,\pi^0 \leftarrow $ from user
\label{line_rec_input_start2}
\While{True}
\label{step_inner_start}
\State $\Omega_{R\pi^+}\leftarrow \cup_{\hat{l} \in \Omega_R} \mbox{arg}\min_{l \in \Omega_{\hat{l}}}c_l+\pi^0_{f_l}-\sum_{u \in N}a_{ul}(\pi^0_u+\nu)$
\label{line_Gen_Q}
\State $\Delta^+_u, \leftarrow \pi^0_u+\nu$ for all $u \in N$
\label{set_delta_1}
\State $\Delta^-_u,\leftarrow \max(0,\pi^0_u -\nu)$ for all $u \in N$
\label{set_delta_2}
 \State $\bar{\pi},\theta \leftarrow $ Solve  \eqref{primal_box} in primal/dual form over $\Omega_{R\pi^+}$
 \If {$\ell^{\Omega^+_R}_{\bar{\pi}}\leq \ell^{\Omega^+_R}_{\pi^{0}}$}
 \label{checkCondStart}
 \State Break
 \label{BreakLine}

 \EndIf
 \label{checkCondEnd}
 \label{line_gen_Dual}
 \State $\vec{\pi} \leftarrow \bar{\pi}-\pi^{0}$
\State $m \leftarrow \min_{\substack{z \in N\cup F\\ \vec{\pi}_z<0}}\frac{-\vec{\pi}_z}{\pi^{0}_z}$ Compute max feasible step size.
\label{compute_step_size}

 \label{compute_Direction}
\State $\eta \leftarrow \max_{\eta \in [\frac{1}{m},1]}\phi_{\eta}$ via Alg \ref{searchProc} 
\label{compute_step_length}
\State $\pi^0 \leftarrow \pi^{0}+\eta m\vec{\pi}$
\label{update_pi_1}
 \EndWhile
 \label{step_inner_end}
 \State Return $\bar{\pi},\theta$
 \label{return_pi}
\end{algorithmic}
\label{ezVer_master}
\end{algorithm}

    \begin{enumerate}
    \item Line \ref{line_Gen_Q}:  Grab $\Omega_{R\pi^+}$.  Here $f_l$ refers to the facility associated with column $l$ meaning $f_l \leftarrow \mbox{arg}\max_{f \in F}a_{fl}$ 
    \item Line \ref{set_delta_1}-\ref{set_delta_2}:   Determine $\Delta^+,\Delta^-$ terms
    \item Line \ref{line_gen_Dual}:  Solve \eqref{primal_box} providing optimal primal/dual solution pair over the box described by $\Delta$.
    \item Line \ref{checkCondStart}-\ref{checkCondEnd}:  If $\bar{\pi}$ is associated with inferior approximate Lagrangian relaxation relative to $\pi^0$  we break and return $\bar{\pi}$.  This assures that a new column is generated very close to the current incumbent dual maximizing solution.   Note that $\bar{\pi}$ corresponds to the minimum possible step-size.
    \item Line \ref{compute_step_size}:  Compute maximum feasible step size as described by \eqref{maxMval}.  If no components appear under the min then this is unbounded and use a large positive number. 
    \item Line \ref{compute_Direction}: Compute the point $\hat{\pi}$ corresponding to traveling the maximum possible distance in the vector corresponding to starting at $\pi^0$ and traveling towards $\pi$. 
    \item Line \ref{compute_step_length}:  Compute the optimal travel distance using Alg \ref{searchProc}. Note that we need only search range $\eta \in [\frac{1}{m},1]$ since the minimum step size is $\frac{1}{m}$.
    \item Line \ref{update_pi_1}:  Update $\pi^0$ to be optima   
    \end{enumerate}
    \item Line \ref{return_pi}:  Return approximate solution to FRMP. 
\end{itemize}

\subsection{Proof of Convergence}
\label{convergAnal}
We now show that at termination of Alg \ref{ezVer} that the solution $\theta$ is optimal for the dual master problem.  We do this as follows. 

Via  \eqref{boundRQ} we know that $c_{l}+\pi_{f_l}+\sum_{u \in N}a_{ul}\pi_u\geq 0$ for each $l \in \Omega_R$.  Thus pricing  can never generate a column already in $\Omega_R$.  

 If Line \ref{checkCondStart} terminates with an inequality then there must be a column with negative reduced cost and hence Alg \ref{ezVer} does not terminate that iteration. If Line \ref{checkCondStart} terminates with an equality then the solution to the FRMP is solved exactly over $\Omega^+_R$ and thus the objective is less than than the master problem. Hence a column with negative reduced cost will be generated if the FRMP objective is greater than the master problem objective.
 \subsection{Limiting the number of inner loop iterations}
 \label{speedInner}
 In this section we accelerate optimization by not solving the FRMP exactly in Alg \ref{ezVer_master} when to do so would require a large number of iterations. We  only run the loop solving the FRMP in Alg \ref{ezVer_master} (Lines:\ref{step_inner_start}-\ref{step_inner_end}) up to a finite number of iterations; where the number of iterations is a fixed user defined parameter. If Lines: \ref{step_inner_start}-\ref{step_inner_end} do not terminate in that period via Line \ref{BreakLine} we do not  terminate optimization  Alg \ref{ezVer} if no negative reduced cost column is found in Line \ref{line_outer_end} but instead continue Alg \ref{ezVer}.  The use of this does not interfere with the convergence guarantees in Section \ref{convergAnal}. In fact it merely consists of generating a column when we have only partially completed optimization over the FRMP and FRMP optimization continues unmodified if no column with negative reduced cost is identified.
\section{Experimental Analysis on the SSCFLP}
\label{exper}
\subsection{Column Projection}
\label{applicationSSCFLP}

Family CG requires in Algorithm \ref{ezVer_master} that we can solve \eqref{projection_eq}. We now show that this is easy for our application of SSCFLP. We write \eqref{projection_eq} for SSCFLP below.  

\begin{subequations}
\label{pricingKnap2}
\begin{align}
\min_{x_u \in \{0,1\} \quad \forall u \in N}c_f+\pi_f+\sum_{u \in N}x_u(c_{fu}-\pi_u)\\
\sum_{u \in N}x_u d_u \leq K_f \label{enfkap2}
\\
x_u=0 \quad \forall u \notin N_l
\label{enfBinary}
\end{align}
\end{subequations}
Observe that any binary valued solution satisfying \eqref{enfBinary} satisfies \eqref{enfkap2} and thus \eqref{enfkap2} can be ignored. Thus an optimizer of \eqref{pricingKnap2} is written as follows $x_u\leftarrow [u \in N_l][c_{fu}<\pi^+_u]$ for all $u \in N$ where $[]$ is the binary indicator function.
%
\subsection{Numerical Experiments}
\label{}

In this section we provide numerical experiments demonstrating the FRMP's effectiveness in accelerating CG convergence when compared to baseline methods. We apply FRMP to the SSCFLP and evaluate its convergence time and iterations required when solving the linear relaxation. We compare against unstabilized CG and against smoothing, which has shown to offer significant speedups on the SSCFLP. We test on 50 randomly generated instances. In each instance we have 50 facilities and 250 customers. Each facility has a capacity of 150 and a fixed opening cost of 5. Each customer has a demand which is randomly generated uniformly over the set $\{1,2,3,4,5\}$. To generate random customer service costs, each customer along with each facility is randomly given a position uniformly on the unit square. Service cost for each facility to each customer are set as the distance from that facility's position to that particular customer.

For smoothing we initialize our center dual solution $\pi^c$ to 0. We send to pricing a convex combination of our center dual solution and the current dual solution to the RMP $\pi_{pricing}=\lambda\pi^c+(1-\lambda)\pi$. We initially set $\lambda$ to 0 and reduce it by 0.1 every time there is a misprice. Following a misprice, when pricing finally produces a negative reduced cost column, we return $\lambda$ back to its initial value of 0.9 and proceed as before. Runtime and iteration count results for the FRMP and smoothing are shown in Table \ref{table:structured}. We show the number of iterations required and the total time required. As well we show the total time required for only solving the LP in the primal. This is presented since the FRMP requires some notable overhead in primal portion of the algorithm that lies outside solving the linear program. This overhead can be vary depending on problem characteristics or implementation so we provide some perspective on performance when its effects are excluded.

\begin{table}[!hbtp]
	\centering
	\scalebox{0.9}{
	\begin{tabular}{|c|c|c|c|} 
		\hline
		\multirow{1}{*}{ } & \multicolumn{1}{c|}{\bf Unstabilized} & \multicolumn{1}{c|}{\bf Smoothing} & \multicolumn{1}{c|}{\bf Family }\\
		\hline
		mean total iterations & 1736.2 & 465.3 & 175.3  \\ 
		\hline
		median total iterations & 1212.5 & 373.5 & 148.5 \\ 
		\hline
		mean total runtime & 2750.0  & 79.7 & 125.1 \\ 
		\hline
		median total runtime & 521.7 & 67.9 & 78.6 \\
		\hline
		mean total LP runtime & 2735.6 & 76.5 & 40.0 \\ 
		\hline
		median total LP runtime & 513.3  & 65.3 & 24.0\\
		\hline
	\end{tabular}}
	\caption{SSCFLP results.}
	\label{table:structured}
\end{table}

\begin{figure}[!hbtp]
	\includegraphics[width=0.49\linewidth]{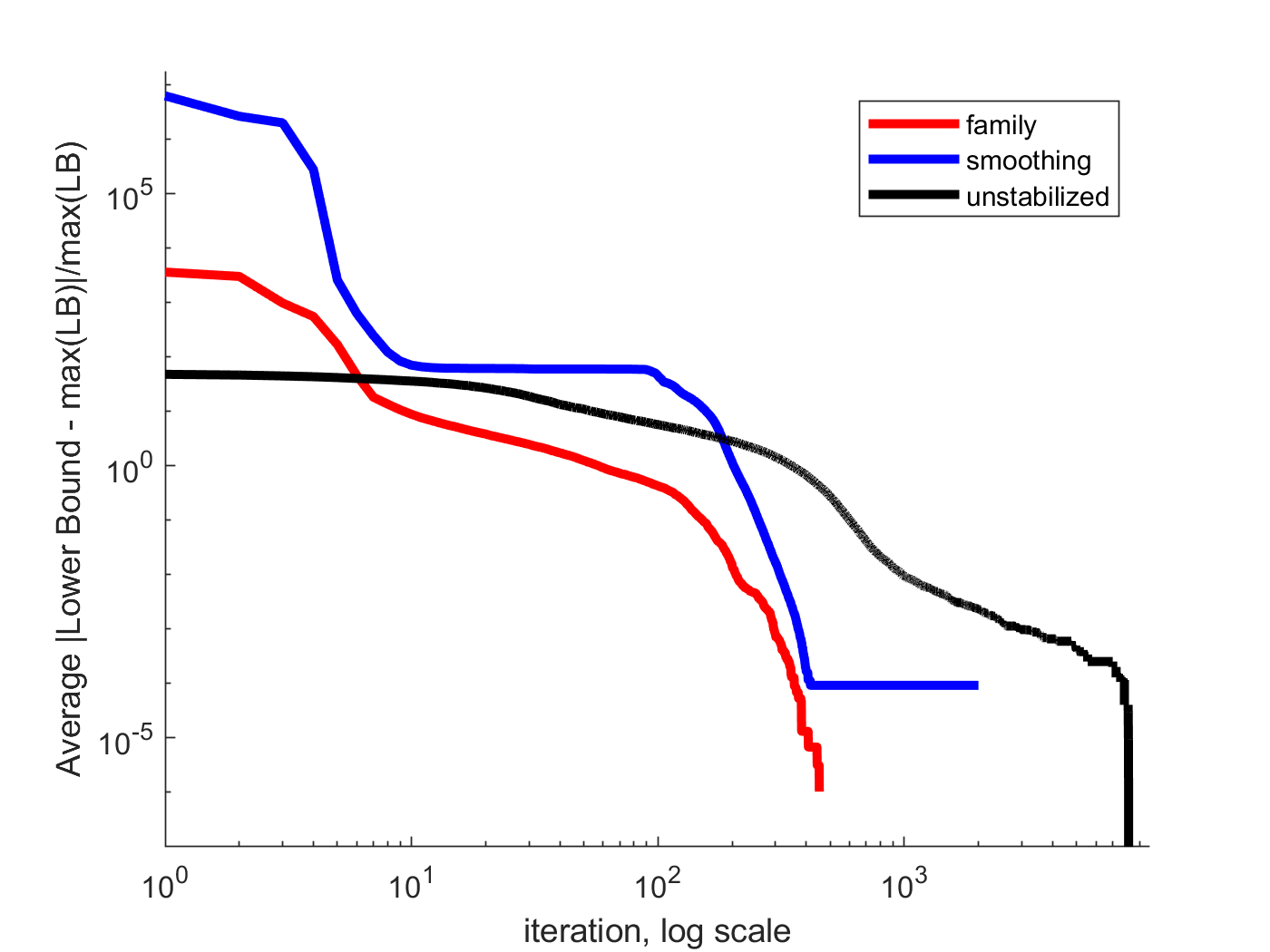}
	\includegraphics[width=0.49\linewidth]{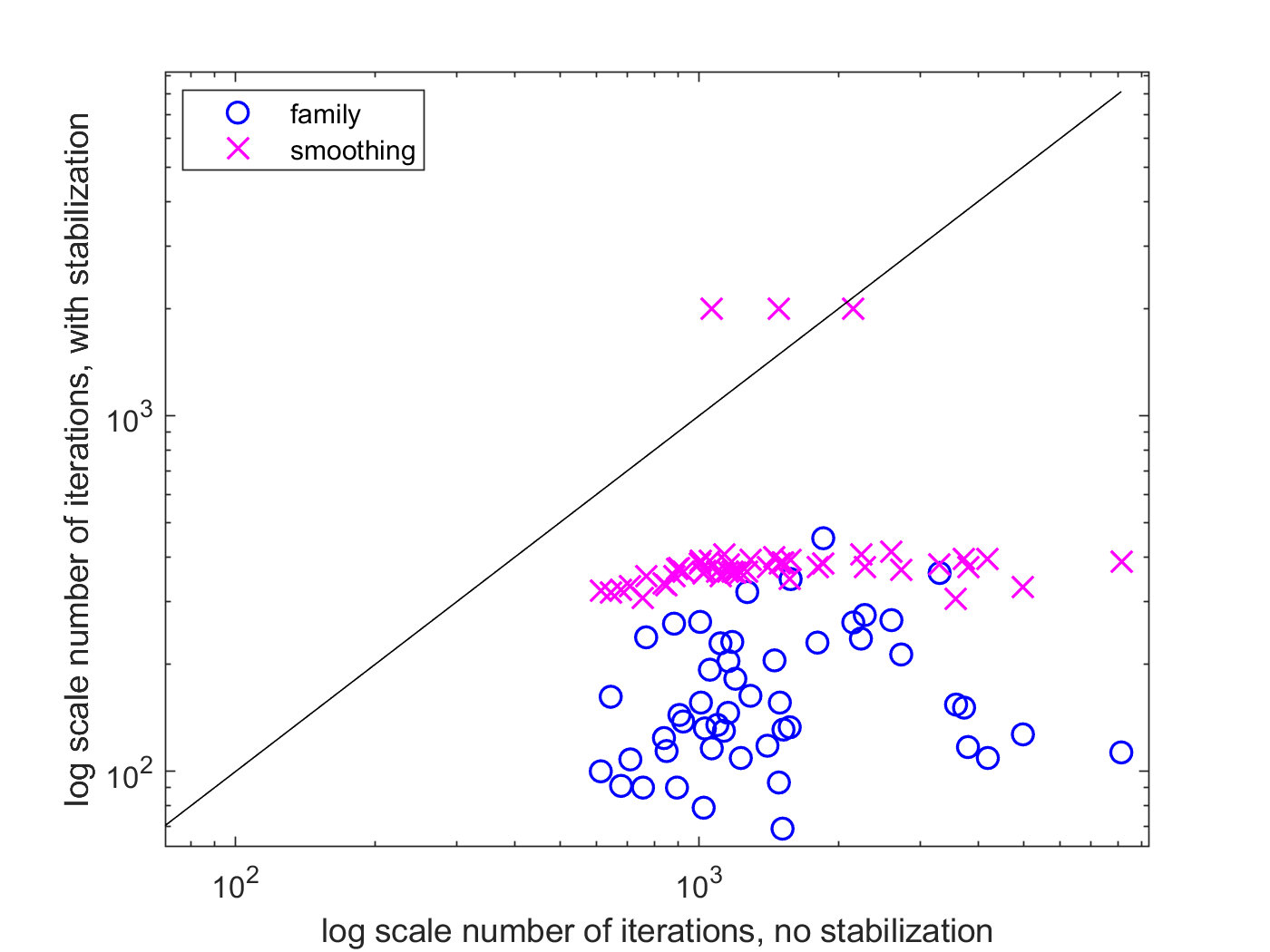}
	\caption{Aggregate results as a function of iteration count.
		\textbf{(Left):} Average relative gap for the lower bound as a function of iteration.  
		\textbf{(Right):} Iteration count comparison of Family RMP vs smoothing over 50 instances.}
	\label{fig:iter}
\end{figure}

\begin{figure}[!hbtp]
	\includegraphics[width=0.49\linewidth]{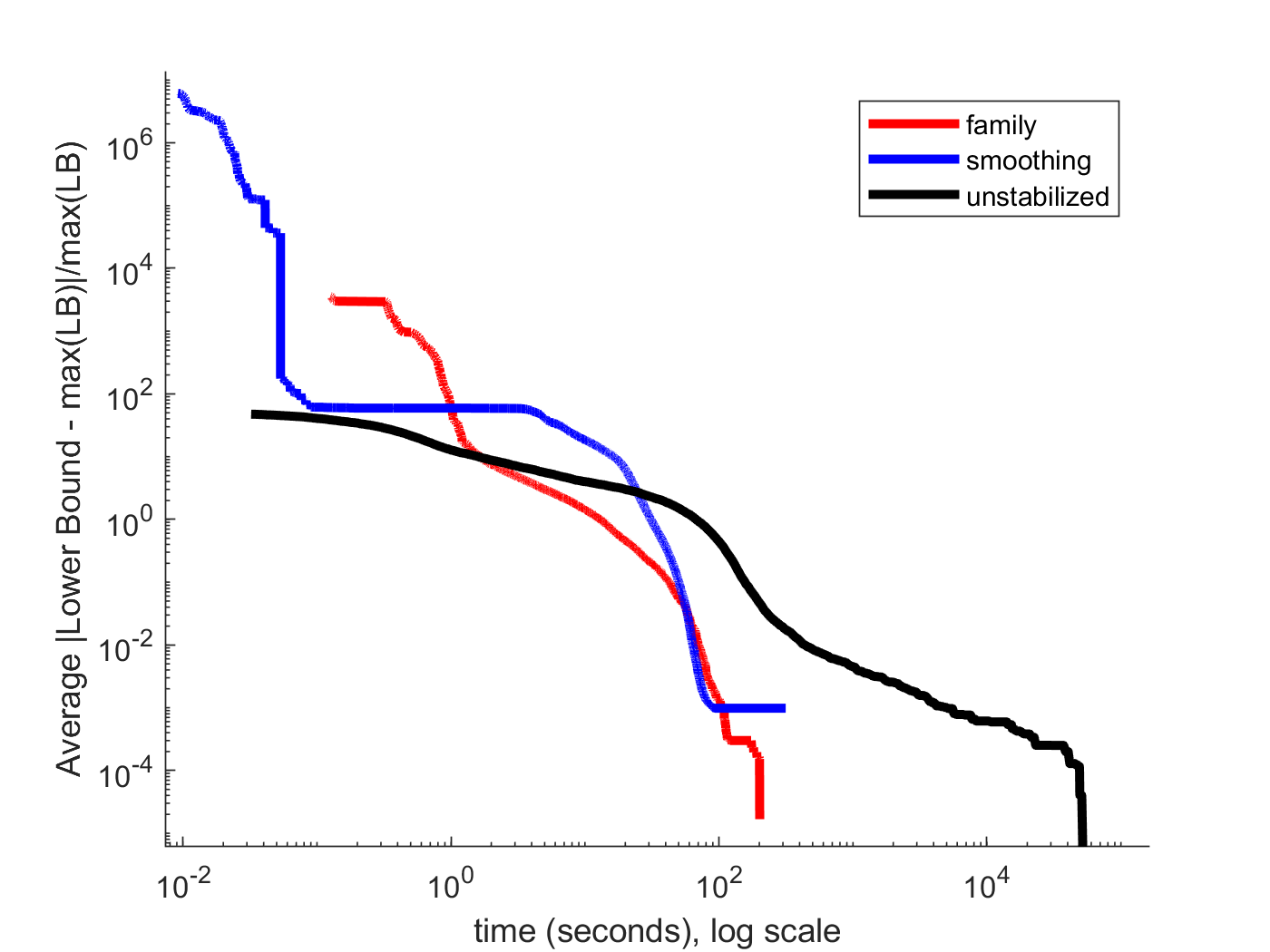}
	\includegraphics[width=0.49\linewidth]{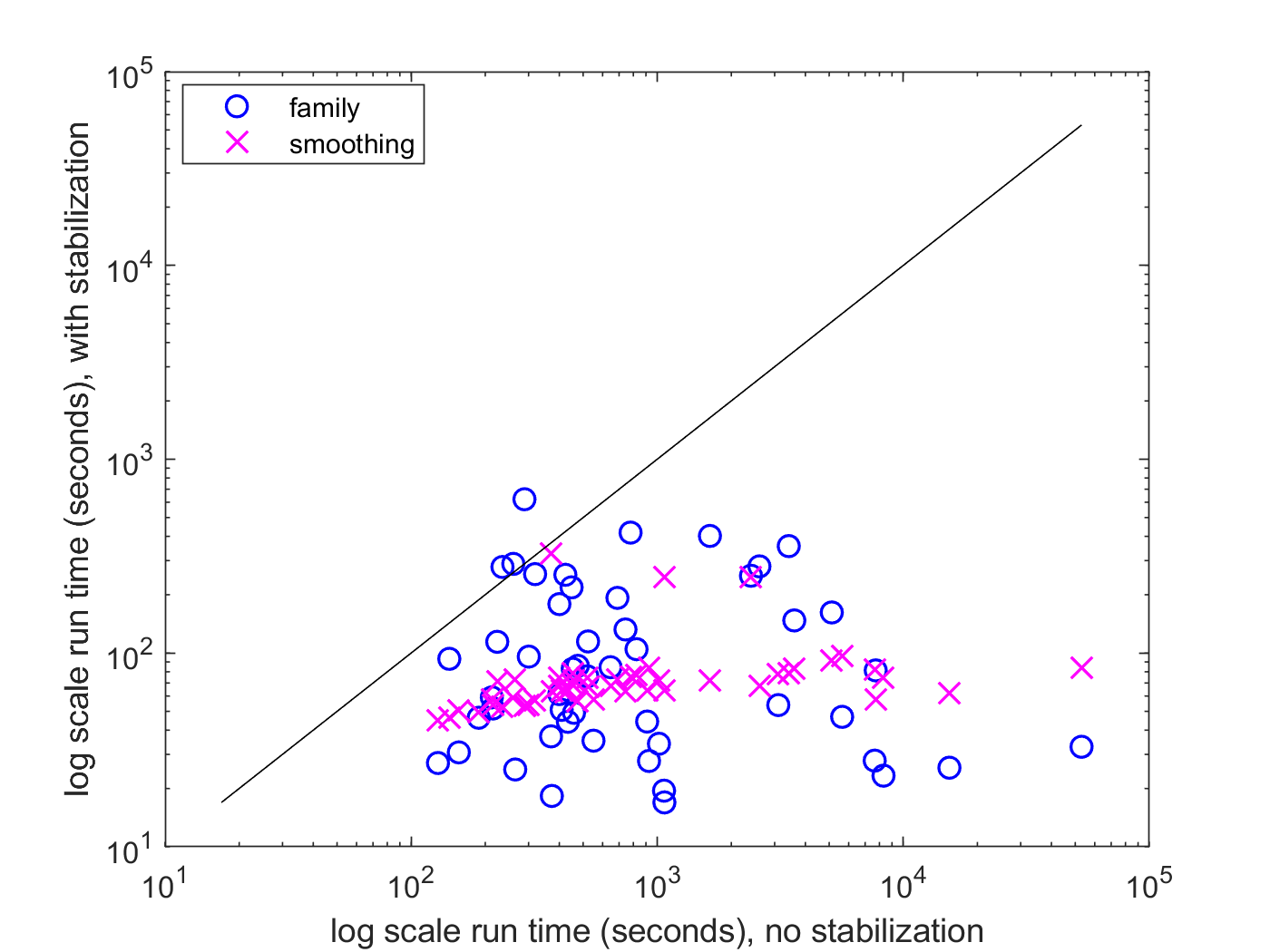}
	\caption{Aggregate results as a function of total runtime.
		\textbf{(Left):} Average relative gap for the lower bound as a function of total runtime.  
		\textbf{(Right):} Total runtime comparison of Family RMP vs smoothing over 50 instances.}
	\label{fig:time}
\end{figure}

\begin{figure}[!hbtp]
	\includegraphics[width=0.49\linewidth]{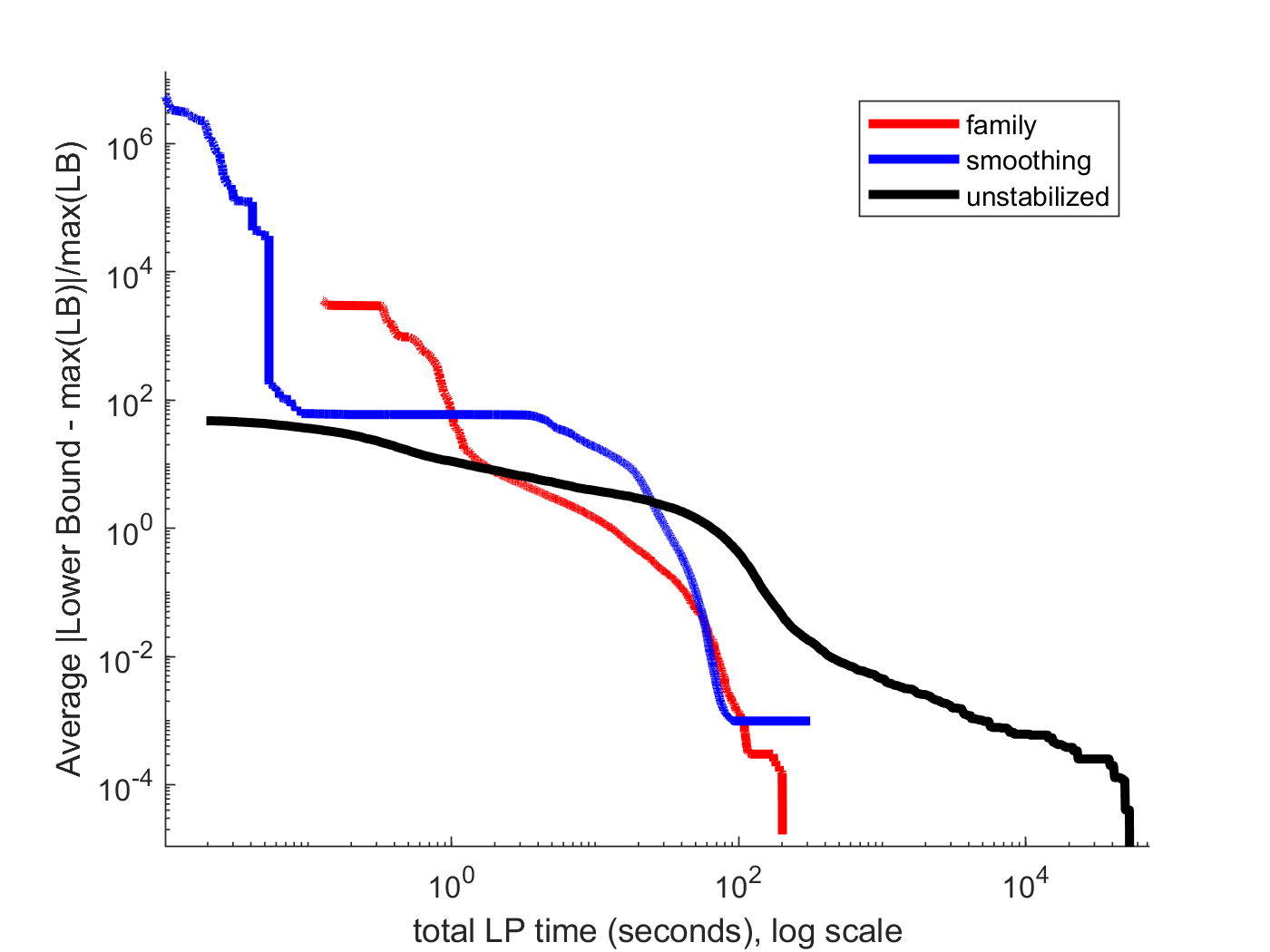}
	\includegraphics[width=0.49\linewidth]{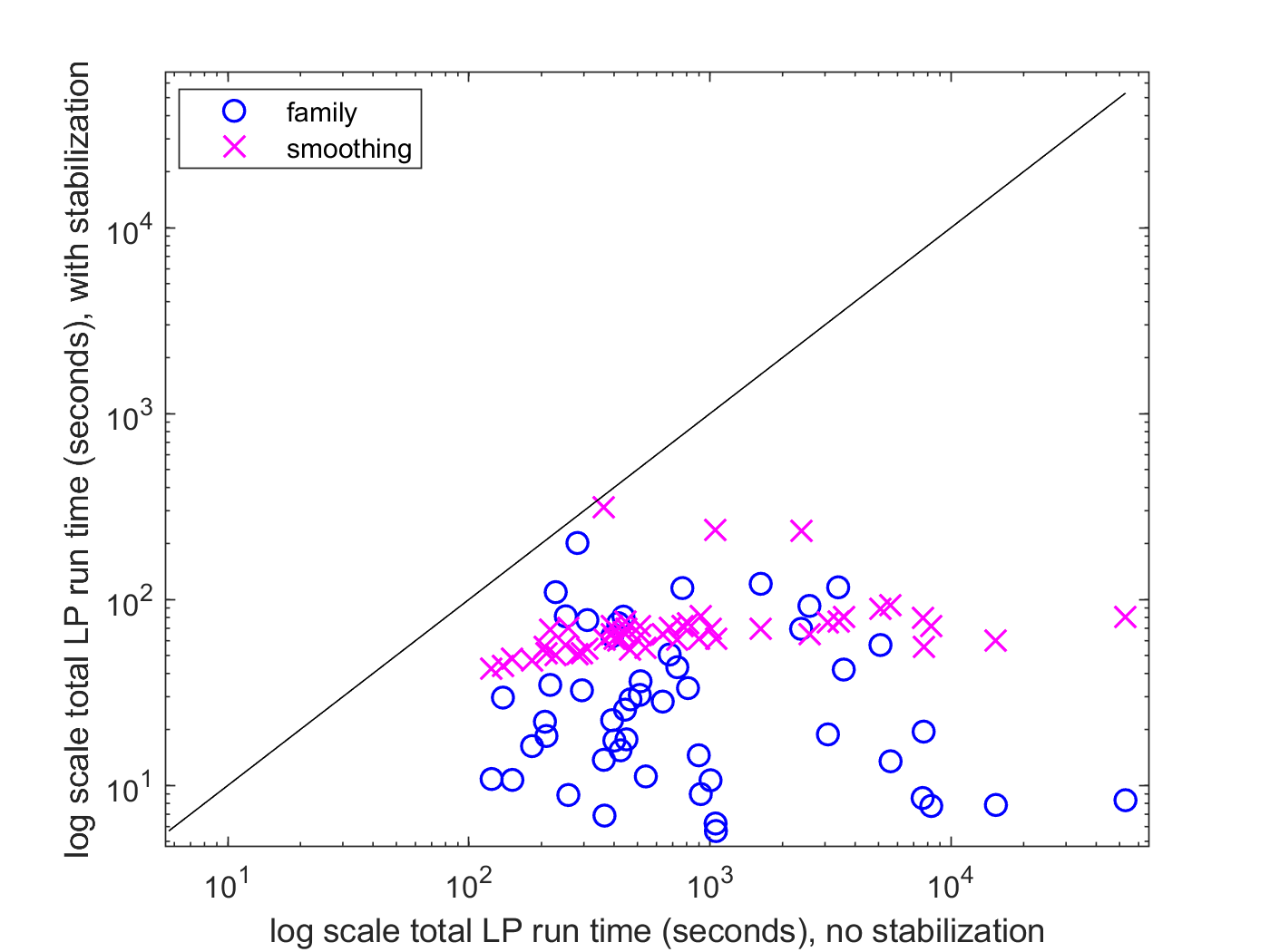}
	\caption{Aggregate results as a function of total LP time.
		\textbf{(Left):} Average relative gap for the lower bound as a function of total LP time.  
		\textbf{(Right):} Total LP time comparison of Family RMP vs smoothing over 50 instances.}
	\label{fig:LPtime}
\end{figure}

The FRMP shows significant reductions in the average number of iterations required (and consequently the number of calls to pricing) when compared to smoothing. Relevant plots for iteration counts are shown in Figure \ref{fig:iter}. When considering total runtime, the FRMP falls short of outperforming smoothing on average. Plots for total runtime are shown in Figure \ref{fig:time}. Looking only at total LP solver time, the FRMP again offers vast improvements over smoothing. Relevant plots for LP runtime are shown in Figure \ref{fig:LPtime}.

\newpage
\section{Future Research}
\label{FutureWork}

First we describe an enhancement to our Family RMP method. Then we outline future analysis of these methods on other important combinatorial optimization problems.   

\subsection{Alternative Families}
In this case we will define the family $\Omega^+_R$ as the same begin $\cup_{l \in R}\Omega_{\hat{l}}$.  However we will change the definition of $\Omega_{\hat{l}}$ to be a superset of what it was before.  The only requirement is that we can solve for the lowest reduced cost member of $\Omega_{\hat{l}}$ easily. 

For the case of SSCFLP we write this set as follows.  Let $a_{dl}$ be the number of items in $l$ containing at least $d$ unit of demand meaning $a_{dl}=\sum_{u \in N;d\leq d_u}a_{ul}$. We define the $\Omega_{\hat{l}}$ to be the set of columns\\ $\hat{l} \in \Omega $  $a_{d\hat{l}}\leq a_{dl}$ for all $d \in [0,1,...,\max_{u \in N}d_u]$.

We write projection pricing below given column $\hat{l}$ associated with facility $f$

$\min_{l \in \Omega_{\hat{l}}}c_l+\pi_f-\sum_{u \in N}a_{ul}\pi_u$

We write this as an optimization problem as follows.  
\begin{subequations}
\label{priceEZ}
\begin{align}
    \min_{x \in \{0,1\} }c_f+\pi_f+\sum_{u \in N}(c_{fu}-\pi_u)x_u
    \sum_{u \in N}x_ud_u\leq K_f\\
    \sum_{u \in N}x_u [d_u\geq d]\leq a_{dl}
\end{align}

\end{subequations}
Observe that we can solve \eqref{priceEZ} by first sorting $(c_{fu}-\pi_u)$ from order of smallest to largest and excluding the positive values. Let $u_i$ be the i'th item in the list .  We can greedily and optimally construct the solution to \eqref{priceEZ} as follows.  
\begin{align}
    x_{u_i} \leftarrow 1 \quad IFF \quad
    \sum_{j<i}x_{u_j}[d_j\leq d]<a_{dl} \forall d\leq  d_u
\end{align}

We need only one additional tool to apply the technique of Family RMP.  We need to produce $\Omega_{R\pi^+}$. We use upper bound terms for $u \in N_{\hat{l}}$ and lower bound terms for $u \notin N_{\hat{l}}$. The replacement equation for \eqref{projSet} is written below.   

\begin{subequations}
\label{projSet2}
\begin{align}
    \Omega_{R\pi^+}=\cup_{\hat{l} \in \Omega_R}\hat{l}_{\pi^+}\\
    \hat{l}_{\pi^+}= \mbox{arg}\min_{l \in \Omega_{\hat{l}}}c_{l}+\sum_{f \in F}a_{fl}\pi_f-\sum_{u \in N_{\hat{l}}}a_{ul}\pi^+_u-\sum_{u \notin N_{\hat{l}}}a_{ul}\pi^-_u
\end{align}
\end{subequations}
This satisfies the properties discussed in Section \ref{genDirec}.  

\subsection{Additional Problems}
In the future we plan to explore additional problems such as those discussed in  \citep{Pessoa2018Automation}. That paper examines nine classical combinatorial optimization problems. Of the problems discussed in that paper, we are most concerned with capaciated vehicle routing and multi-activity shift scheduling. In addition, we plan to immediately turn our attention to applying these techniques to the various optimization problems arising in automated warehousing operations (see for example \citep{haghani2021multi}). Variations on multi-robot routing and product placement in shelf-to-picker and picker-to shelf problems in these operations are of keen current interest. 


\section{Conclusion}
\label{conc}
We have introduced a new methodology to accelerate the convergence of column generation when applied to set-covering-based formulations by stabilizing dual optimization. Our methodology can be applied on any set cover-CG based formulation and we believe that it will accelerate the adoption of column generation methods for optimization problems that must be solved in real-time. 

Our approach seeks to solve a restricted master problem over the set of all columns in the family of columns in the restricted master problem(RMP).  Here the family of a column  $l$is simply all columns with item set that is a subset of those in $l$ and for which any structural properties are preserved.  In vehicle routing the structure describes the order of the items visited while in facility location it describes the facility associated with the column.  

Our approach is identical to standard CG with regard to pricing though it differs with respect to the solution to the restricted master problem.  Since optimization over the family can not be done explicitly due to combinatorial explosion, a coordinate ascent approach is employed. Directions are generated via solving over an upper bound over a finite box and the optimal amount of travel distance is determined by binary search. The optimization over the finite box uses columns that remove items for which the dual variable range does not justify the inclusion of.

FRMP CG circumvents the difficulties in the box-step method on which it is based. First it circumvents the need to choose a complex schedule for step size $\nu$ by following the direction produced so as to maximize an approximation to the Lagrangian relaxation. Thus a small fixed $\nu$ can be used and large moves in the dual space can still be achieved. The projection step at CG ensures that each column in the RMP is mapped to a column providing a constraint to the dual solution, ensuring that the intermediate dual solutions are not bound by only a small number of columns.  

We demonstrate the efficacy of our approach on the SSCFLP. In future work we intend to explore vehicle routing problems and related multiple robot routing problems. In vehicle routing problems, pricing over the family of a given column computes the lowest reduced cost column containing a subset of the items in column $l$ and obeying the ordering of the items in $l$. Thus pricing involves solving a shortest path problem and not a resource constrained shortest path problem. That difference results in an immense improvement in efficiency. We also intend to explore adaptive step-size changing mechanisms to improve our current approach which uses a fixed step size throughout optimization.  

\bibliographystyle{abbrvnat} 
\singlespacing
\bibliography{family_RMP}
\appendix

\end{document}